\documentclass[aps,prl,twocolumn,a4paper,floatfix,showpacs]{revtex4} 
\usepackage{amsmath}
\usepackage{amsfonts}
\usepackage{epsfig}
\graphicspath{{./}}
\begin{document}
  \title{Probing the FFLO phase by double occupancy modulation spectroscopy}
   \author{Anna Korolyuk} 
   \affiliation{Department of Applied Physics, P.O. Box 5100,
                02015 Aalto University, Finland}

   \author{Francesco Massel}
   \affiliation{Department of Applied Physics, P.O. Box 5100,
                02015 Aalto University, Finland}

  \author{P\"aivi T\"orm\"a}
  \email{paivi.torma@hut.fi}
   \affiliation{Department of Applied Physics, P.O. Box 5100,
                02015 Aalto University, Finland}


  \begin{abstract}
    We propose here that for a spin-imbalanced two-component
    attractive Fermi gas loaded in a 1D optical lattice in presence of
    an harmonic confining potential, the observation of the change in
    the double occupancy after a lattice depth modulation can provide
    clear evidence of the Fulde-Ferrell-Larkin-Ovchinnikov (FFLO)
    phase. Simulating the time evolution of the system, we can
    characterize the double occupancy spectrum for different initial
    conditions, relating its features to the FFLO wavevector $q$. In
    particular, the narrowing of the width of the spectrum can be
    related, through Bethe-ansatz equations in the strongly
    interacting limit, to the FFLO wavevector $q$.
  \end{abstract} 
 
  \maketitle Ultracold atoms trapped in optical lattices have become
  an important tool to mimic strongly correlated condensed matter
  systems, leading to the possibility to explore regimes unattainable
  within the traditional solid state framework.  Recently, a
  considerable experimental effort \cite{Liao2009,Partridge2006} has
  been devoted to the analysis of two-component spin-imbalanced Fermi
  gases.  Theoretical investigations
  \cite{Orso2007,Koponen2007,Feiguin2007,Batrouni2008,Luscher2008,Liu2008,Bakhtiari2008,Paananen2008,Edge2009,Kakashvili2009}
  have revealed that, in the characterization of 1D spin-imbalanced
  Fermi gases, a major role is played by the
  Fulde-Ferrell-Larkin-Ovchinnikov (FFLO) state \cite{Fulde1964}. In a
  solid-state context the FFLO phase has been investigated in heavy
  Fermions systems \cite{Radovan2003}, with techniques ranging from
  heat capacity to nuclear magnetic resonance measurements, even
  though conclusive evidence of its existence is still missing. In
  ultracold gases, even though it has been suggested that its presence
  can be detected through various measurements such as noise
  correlation \cite{Paananen2008,Liu2008,Greiner2005}, radio-frequency
  spectroscopy \cite{Bakhtiari2008}, collective modes analysis
  \cite{Edge2009} , and local density profile measurement
  \cite{Kakashvili2009}, no direct experimental evidence of the FFLO
  phase has been found in these systems. Nevertheless in the experiment
  conducted at Rice University \cite{Liao2009}, the density profile of
  each component has been measured, exhibiting a behavior compatible
  with the theoretical analyzes focusing on the characterization of
  the FFLO phase. Compared to previous theoretical suggestions, our
  approach relies on a simple experimental setup and at the same time
  provides unequivocal signature of the FFLO phase.

  In particular, we propose that a clear experimental evidence of the
  FFLO phase in a 1D optical lattice can be provided by the
  measurement of the double occupancy (d.o.),-- \textit{i.e.} the
  number of sites populated by two atoms-- after a periodic lattice
  modulation of the initial state at different frequencies (d.o.
  modulation spectrum, as proposed in \cite{Kollath2006}).  This
  technique has been employed to observe the appearance of the Mott
  gap in a repulsive two-component Fermi gas \cite{Jordens2008} and it
  has been suggested as a possible tool to detect the
  antiferromagnetic phase in such systems \cite{Sensarma2009,Massel2009}.
  Performing the same kind of experiment for an attractive gas is well
  within reach of the current experimental techniques and, as we will
  show here, can provide clear evidence of the FFLO phase through a
  reduction of the width in the d.o. spectrum directly related to the
  FFLO vector $q$. The underlying physics is simple: the existence of
  a collective momentum $q$ restricts the available momentum states of
  the excitations, thus narrowing the spectrum.
 

  In presence of a parabolic confining potential, we assume that the
  system is described by the Hubbard Hamiltonian
  \begin{equation}
    \label{eq:hubb}
    H=H_{J}+H_{U}+\sum_{i}^{L}V_i \left(n_{i\uparrow}+n_{i\downarrow}\right),
  \end{equation}
  where $H_{J}=-J\sum_{i,\sigma=\uparrow\,\downarrow
  }^{L}c_{i\sigma}^{\dagger}c_{i+1\sigma}+h.c.$,
  $H_{U}=-U\sum_{i}^{L}n_{i\uparrow}n_{i\downarrow}$,
  $V_i=V(i-\frac{L}{2})^{2}$, $J$ is the hopping amplitude, $-U$
  is the on-site attractive interaction, $V$ the global confining
  potential, and the polarization $P$, defined as
  $P=\left(N_{\uparrow}-N_{\downarrow}\right)/\left(N_{\uparrow}+N_{\downarrow}\right)$,
  with $N_\sigma=\sum_\sigma n_{i\sigma}$.

  The lattice depth modulation proposed in
  \cite{Kollath2006,Jordens2008} can be modeled by the modulation of
  the hopping amplitude $J(t)=J+\delta J\cos(\omega t)$
  \cite{Sensarma2009,Massel2009}. Since we are interested in
  excitations which lead to pair breaking, it is natural to focus on
  modulation frequencies close to the energy $U$, related to pairing,
  \textit{i.e.}  we concentrate on the transition between the first
  and the second Hubbard band.  Intuitively, the process we are
  interested in might be understood as the transition between the
  ground state and a state where a pair has been broken by the hopping
  modulation.

\begin{figure}
 \includegraphics[width=0.32\textwidth]{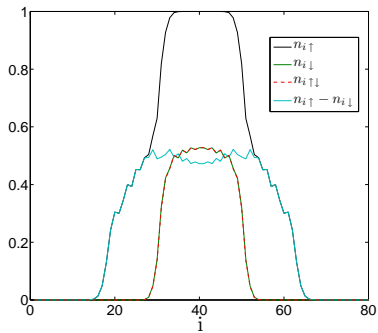}
 \caption{(color online) Particle and pair densities and the
   difference $n_{i\uparrow}-n_{i\downarrow}$ for the polarization
   $P=0.5$ ($N_{\uparrow}=30$,$N_{\uparrow}=10$).}
  \label{fig:GS_imbalanced}
\end{figure}

\begin{figure}
\includegraphics[width=0.35\textwidth]{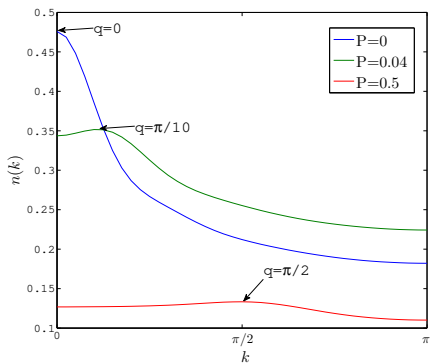}
\caption{(color online) Pair momentum distribution $n(k)$ for the polarizations
  $P=0$, $P=0.04$, $P=0.5$. The maxima are at $q=0$,
  $\simeq \frac{\pi}{10}$,$\simeq \frac{\pi}{2}$, respectively.}
  \label{fig:Pair_momentum_distribution}
\end{figure}

  Our approach to the problem is twofold. We first perform numerical
  simulations of the ground state and of the dynamical evolution of
  the system. We then move to the analysis of the results in terms of
  Bethe-ansatz (BA), in the limit $U/J\to \infty$. The numerical
  simulations, both for the ground-state calculation and for the time
  evolution, are performed with the aid of a time-evolving block
  decimation (TEBD) code \cite{Vidal2003}, which can be regarded as a
  quasi-exact method for the analysis of 1D quantum systems. In the
  spin-polarized case, for the range of parameters that we have
  considered ($U=-10$, $J=1$, $N_{\uparrow}+N_{\downarrow}=40$, $P\geq
  0.04$, $V=0.005$) the ground state consists of a central region
  where $\left\langle n_{i\uparrow} \right\rangle>\left\langle
    n_{i\downarrow}\right\rangle>0$ and an outer, fully polarized
  region ($\left\langle n_{i\uparrow} \right\rangle>0$, $\left\langle
    n_{i\downarrow} \right\rangle=0$). Moreover, in the central region
  of the trap $\left\langle n_{i\downarrow}\right\rangle \simeq
  \left\langle n_{i\uparrow} n_{i\downarrow}\right\rangle$ implying
  that due to the strong interaction considered here all minority
  particles $n_{i\,\downarrow}$ are paired (Fig.
  \ref{fig:GS_imbalanced}).  The periodic spatial dependence of
  $n_{i\uparrow}-n_{i\downarrow}$ suggests the presence of the FFLO
  state \cite{Tezuka2008}.
In order to give a quantitative estimate, we extract the
value of the FFLO wavevector $q$ from the pair correlation function
$\langle
c_{j\uparrow}^{\dagger}c_{j\downarrow}^{\dagger}c_{i\uparrow}c_{i\downarrow}\rangle$
and its Fourier transform $n_{pair}(k)$, defining $q$ as the maximum
in the distribution $n_{pair}(k)$
(Fig. \ref{fig:Pair_momentum_distribution}).

\begin{figure}
\includegraphics[width=0.32\textwidth]{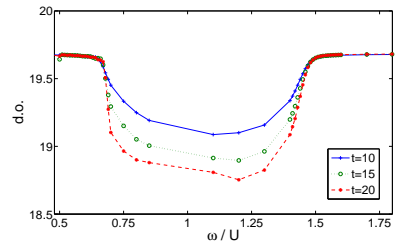}
\caption{(color online) Double occupancy $\overline{D}_{\omega}(t)$ as
  function of frequency $\omega$ for times $t=10,  15, 20$ for the
  balanced case $N_{\uparrow}=N_{\downarrow}=20$.}
\label{fig:Do_as_function_of_frequency}
\end{figure}

We first examine the properties of the system for $P=0$. After
calculating the ground state for the Hamiltonian, we turn on the
modulation of the hopping amplitude. At each timestep we calculate the
total d.o. spectrum $D_{\omega}(t)=\sum_{i=1}^{L}\left\langle
n_{i\uparrow}n_{i\downarrow}\right\rangle $, where the modulation
frequency $\omega$ is centered around the value of the interaction
strength $\left|U\right|$.
In Fig.  \ref{fig:Do_as_function_of_frequency}, the d.o. spectrum
$\overline{D}_{\omega}(t)\mid_{t=50}$ is plotted for frequencies
$\omega\in\left[0.5;1.8\right]$, where
$\overline{D}_{\omega}(t)\mid_{t=50}$ is the average between the local
maxima and minima in the small-time dynamics of the d.o.. The
spectrum shows a band-like structure with $\omega_{min}\simeq 0.68$
and $\omega_{max}\simeq 1.5$.  As we will later show, the band in Fig.
\ref{fig:Do_as_function_of_frequency} can be explained in terms of the
excited states within the second Hubbard band.

\begin{figure}
\includegraphics[width=0.35\textwidth]{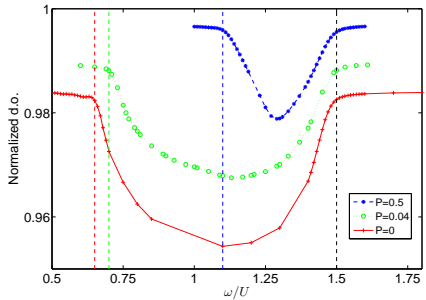}
\caption{(color online) Normalized double occupancy
  $\overline{D}_{\omega}(t)/N_{\downarrow}$ as a function of frequency
  $\omega$ at $t=10$ in a trap for three cases with polarizations
  $P=0$, $0.04$, $0.5$ corresponding to $N_{\downarrow}=20$, $22$,
  $10$. The vertical lines correspond to $E_{high}$ (black line), and
  to $E_{low}$ for different polarizations. As mentioned in the text,
  the (normalized) local minimum for $\overline{D}_{\omega}(t)$
  decreases for increasing number of pairs. Specifically it is located at
  $19.66$, $21.76$, $9.96$ for $N_{\downarrow}=20$, $22$, $10$
  respectively.}
 \label{fig:Do for different polarisation}
\end{figure}

We will now turn our attention to the numerical results for the double
occupancy spectrum of the spin-polarized case (Fig.\ref{fig:Do for
  different polarisation}).
The first important aspect is the decrease in the reduction of the
d.o.  as the number of paired particles is increased (see Fig.
\ref{fig:Do for different polarisation}). This feature can be easily
understood considering that the lattice modulation at frequencies
close to $U$ affects the paired component of the gas only and hence
the number of broken pairs is reduced accordingly.  However, the most
prominent feature of the spectrum in the spin polarized case is the
reduction of the width of the band.  In particular, while the position
of its upper limit is independent of the polarization, the lower limit
depends strongly on $P$. The main goal of our analysis is to show that
the width of the band $\Delta \omega$ can be described by the relation
$\frac{\Delta\omega}{U}=\frac{4J}{U}(1+\cos q)$, where $q$ is the FFLO
wavevector, calculated from the ground-state value of $n_{pair}(k)$.
We thus claim that the determination of the d.o.  modulation spectrum
in an imbalanced gas allows the direct determination of the $q$ vector
characteristic of the FFLO phase.

The physical situation depicted here can be analyzed in terms of the
mapping between the attractive and the repulsive Hubbard model.
Changing $U \to -U$, the single-site basis states can be mapped
according to the following scheme
\begin{equation}
  \label{eq:map}
  |\uparrow \downarrow \rangle \leftrightarrow  |\uparrow \rangle, \quad
  |\emptyset \rangle \leftrightarrow  |\downarrow \rangle. 
\end{equation}  
For repulsive interaction, the hopping modulation results in an
increase of the d.o., since, in that case, the modulation cause the
formation of a doubly occupied and an empty site \cite{Sensarma2009,Massel2009}. In
the case analyzed here the opposite process takes place: a doubly
occupied/empty site ``pair'' is broken. However, as a consequence of
the mapping, the bandwidth for the two processes is the same.  



In order to explain the results obtained we will consider here the BA
solution for the open-boundary conditions (OBC) Fermi-Hubbard model in
the limit $U/J \to \infty$. 
In the case $U>0$, it is possible to prove that
the excitations of the system can be described in terms of
$N=N_\uparrow+N_\downarrow$ spinless fermions with energy and momenta
given respectively by 
\begin{align}
  \label{eq:e_k}
  E&=-2J\sum_{j=1}^{N}\cos k_{j}, \quad  \nonumber\\
  k_{j}&=\frac{\pi}{L+1}I_{j} \quad I_{j}\in\mathbb{N} \textrm{,}\,
  j=\left[1\ldots N\right],
\end{align}
where Eq. \eqref{eq:e_k} can be directly obtained from the $U/J\to
\infty$ limit of the BA equations (see Supplementary Information).
The distribution of $I_{j}$ should correspond to a condition where
energy is minimized. In the half-filled case, the energy minimization
condition is given by $I_{j}=\left[1\ldots L\right]$, leading to
$E=-2J\sum_{j=1}^{L}\cos k_{j}=0$, $p=\sum_{j=1}^{L}k_{j}$.  

We now turn to the analysis of the attractive interaction case.
Following the mapping described in Eq. \eqref{eq:map}, the total
number of up spins $N_{\uparrow}$ in the repulsive case maps to the
total number of pairs $N_{\uparrow\downarrow}$ and $N_{\downarrow}$ to
the number of empty sites $N_{\emptyset}$, leading to
$N=N_{\emptyset}+N_{\uparrow\downarrow}$. In the strongly attractive
regime, we can assume that all down particles are paired, leading to
$N_{\uparrow\downarrow}=N_{\downarrow}$.  $N_{\emptyset}$ is the
number of sites which are neither occupied by a pair
($N_{\uparrow\downarrow}=N_{\downarrow}$) or by an unpaired majority
atom ($N_{\uparrow}-N_{\uparrow\downarrow}$), and hence
$N=L-(N_{\uparrow}-N_{\downarrow})$, leading to

\begin{equation}
  \label{eq:e_k_fflo}
  E=-2J\sum_{j=1}^{N_{\downarrow}-N_{\uparrow}}\cos k_{j},\quad
  k=\sum_{j=1}^{N_{\downarrow}-N_{\uparrow}}k_{j}.
\end{equation}
From Eq. \eqref{eq:e_k_fflo}, it is possible to relate the Fermi
momentum for the spinless Fermion gas to the polarization $P$, namely
consider $ k_F=\pi(N_\uparrow-N_\downarrow)/(L+1)$ and then observe
that the FFLO momentum, defined as $q=\pi \rho P$ with
$\rho=(N_\uparrow+N_\downarrow)/L$, coincides with $k_F$ (if $L\simeq
L+1$). Obviously, for a half-filled system $N_\uparrow+N_\downarrow=L$
and hence $k_F\simeq \pi P=q$.  Even the situation where a parabolic
confining potential is present can be intuitively understood in terms
of an ``effective'' FFLO vector, determined by the spatially dependent
density of the system (see \cite{Luscher2008}).

\begin{figure}
\includegraphics[width=0.3\textwidth]{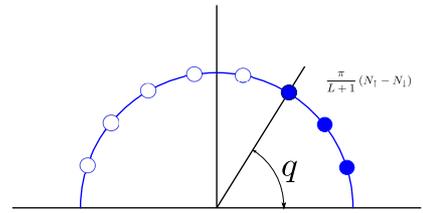}
\caption{Schematic representation of the Fermi sea for the spinless
  Fermions, the Fermi wavevector lies at $q$}
\label{fig:Spinless fermions}
\end{figure}

The effect of the hopping modulation is to create two
\textit{fermionic} excitations (corresponding to the up and down
fermions originating from the breaking of the pair) above the Fermi
energy of the spinless Fermions given in Eq. \eqref{eq:e_k_fflo}. The
change of the kinetic energy imposed by the presence of these two excitations
with respect to the ground state is given by
\begin{equation}
  \label{eq:e_kin}
  \Delta E_{kin}=-2J(\cos k_{1}+\cos k_{2})
\end{equation}
with $-1\leq\cos k_{1,2}<\cos q$. In addition to $\Delta E_{kin}$, the
pair breaking also involves a change in the interaction energy $\Delta
E_{int}=U$.  The total energy difference associated to the breaking of
the pair can be thus expressed as $\Delta E=-2J(\cos k_{1}+\cos
k_{2})+U$.  We then expect that in a d.o. modulation experiment for an
imbalanced gas the pair breaking band will lie between
$E_{low}=U-4J\cos q$ ($k_1=q$, $k_2=q$ ) and $E_{high}=U+4J$
($k_1=\pi$, $k_2=\pi$), leading to a bandwidth
\begin{equation}
  \label{eq:bwidth}
  \frac{\Delta \omega}{U}=\frac{4J}{U}(1+\cos q).  
\end{equation}
In addition, the finite value of $U/J$ in the numerical simulations
implies a shift in the d.o. spectrum $U \to U^*$ leading to
\begin{equation}
 \label{eq:e_tot}
  E_{low}=U^*-4J\cos q,\quad  E_{high}=U^*+4J,
\end{equation}
keeping the value of $\Delta \omega$ unchanged. This shift is
connected to the shift of the ground-state energy within the first
Hubbard band induced by the finite value of the ratio $U/J$. The
explicit calculation of the Hubbard spectrum for a two-site system
allows to get a qualitative understanding of the physical reason
behind this phenomenon. More specifically, the finite value of $U/J$
implies a lowering of the ground-state energy with respect to the case
$U/J\to \infty$, along with a removal of the degeneracy connected to
the spin degree of freedom (see Supplementary Information).
The relation given by Eq.\eqref{eq:bwidth} as well as the
values $E_{low}$ and $E_{high}$ are nevertheless still valid. In Fig.
\ref{fig:Do for different polarisation} it is possible to observe how
the numerical results correspond to our analytical description.

\begin{figure}
\includegraphics[width=0.45\textwidth]{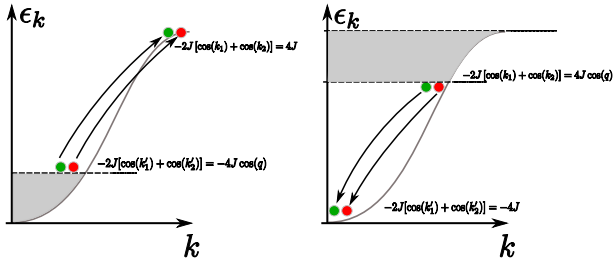}
\caption{(left) Representation of the scattering process between two
  particles with initial momentum $k_1=q$ and $k_2=q$ and final
  momentum $k'_1=\pi$ and $k'_2=\pi$, corresponding to the kinetic
  energy transfer $\Delta E_{max}=4J+4J\cos q$ . The pair momentum $q$
  restricts the available initial states. (right) Scattering process
  between the states $k_1=\pi+q$, $k_2=-\pi+q$ and $k'_1=0$ and
  $k'_2=0$. }
\label{fig:scatter}
\end{figure}

Intuitive understanding of the BA results can be provided by
considering a related example, namely the inelastic scattering of
particles (Fig.  \ref{fig:scatter}). A restriction imposed on the
possible values of the momenta --in our case dictated by the FFLO
wavevector $q$ of the initial state-- implies a reduction of the
bandwidth associated with the scattering process. If the particles
considered have initial momenta and final momenta $k_{1}$, $k_{2}$ and
$k'_{1}$, $k'_{2}$ respectively, the maximum kinetic energy change in
the scattering process will be (assuming the lattice dispersion
relation) $\Delta E_{max}=4J+4J\cos q$ (for
$k_1=q,\,k_2=q,\,k'_1=\pi,\,k'_2=\pi$), and the minimum will be given
by $\Delta E_{min}=-4J-4J\cos(q)$ (for
$k_1=\pi+q,\,k_2=-\pi+q,\,k'_1=0,\,k_2'=0$), hence the largest possible
difference in the kinetic energy associated to the scattering process
will be given by $\Delta E_{max}-\Delta E_{min}=8J(1+\cos q)$. This
simple example illustrates how, in general, the limitation of the
accessible momentum states is reflected in the reduction of the
bandwidth.

Since we are addressing a possible experimental setup to detect the
FFLO phase in 1D, it is necessary to address the role of temperature.
In \cite{Liu2008,Casula2008}, the temperature stability of the FFLO in
1D traps has been considered. In particular in \cite{Liu2008} the
transition temperature $T_c$ between a phase-separated FFLO+normal
$\to$ normal phase is discussed, leading to $T_c\simeq 0.2 T_F$. This
result is obtained within a mean-field picture, providing an
approximate upper limit of the temperatures needed to observe the FFLO
phase. This range of temperature seem to be well within reach in
present experiments. In \cite{Liao2009} temperatures $\simeq 0.1 T_F$
have been reported, suggesting that the FFLO phase could be observed in
the near future.

Through a combination of numerical simulations and analytical results
expressed in terms of BA equations, we have been able to relate the
d.o. modulation spectrum to the presence of a FFLO state, giving a
quantitative estimate of the bandwidth narrowing in terms of the
wavevector $q$. Our analysis establishes the first simple clear
experimental tool to detect and quantitatively characterize the FFLO
phase in ultracold gases in quasi 1D optical lattices. It also shows,
on more general grounds, how a collective (pair) momentum can be
related to observable quantities in 1D systems in a simple and clear
manner.

This work was supported by the National Graduate School in Materials
Physics and Academy of Finland (Project No. 213362, No. 217045, No.
217041, No. 217043), and conducted as a part of a EURYI scheme grant,
see www.esf.org/euryi. We acknowledge CSC -- IT Center for Science
Ltd. for the allocation of computational resources.


\end{document}